# Magnetic-field-induced finite-size effect in the high-temperature superconductor YBa$_2$Cu$_3$O$_{7-\delta}$: a comparison to rotating superfluid $^4$He

R. Lortz [1,2], C. Meingast [1], A. I. Rykov [3], and S. Tajima [3]

[1]*Forschungszentrum Karlsruhe, Institut für Festkörperphysik, 76021 Karlsruhe, Germany*
[2]*Fakultät für Physik, Universität Karlsruhe, 76131 Karlsruhe, Germany*
[3]*Superconductivity Research Laboratory-ISTEC, 10-13 Shinonome I-Chome, Koto-ku, Tokyo 135, Japan*

The effect of strong magnetic fields (11 T) on superconductivity in YBa$_2$Cu$_3$O$_{7-\delta}$ is investigated using high-resolution thermal expansion. We show that the field-induced broadening of the superconducting transition is due to a finite-size effect resulting from the field-induced vortex-vortex length scale. The physics of this broadening has recently been elucidated for the closely related case of rotating superfluid $^4$He [R. Haussmann, Phys. Rev. B60, 12373 (1999)]. Our results imply that the primary effect of magnetic fields of the order of 10 T is to destroy the phase coherence; the pairing, on the other hand, appears to be quite insensitive to these fields.

In classical mean-field type-II superconductors there exists a well-defined upper critical field, H$_{c2}$(T), above which superconductivity is fully destroyed by the applied magnetic field. The value of H$_{c2}$, which is an important fundamental parameter of a superconductor and provides a measure of both the Ginzburg-Landau coherence length ξ$_{GL}$ and the strength of the pairing, can reliably be determined by for example resistive, magnetic or calorimetric experiments. In contrast, in high-temperature superconductors (HTSC) the superconducting transition is significantly broadened by a magnetic field, hindering a straightforward determination of H$_{c2}$ and/or ξ$_{GL}$. The physics of this broadening has received considerable experimental [1-4] and theoretical [5,6] attention, and experimental data have been controversially analyzed using the scaling approaches of the 3d-XY and/or lowest-Landau-level (LLL) fluctuation models.

In a recent theoretical paper [7], Haussmann showed that a similar broadening is expected for rotating superfluid $^4$He, in which the rotational frequency is analogous to the magnetic field in a superconductor. A simple explanation of this broadening was given in terms of a finite-size effect due to the additional vortex-vortex length scale, which prevents the correlation length from diverging resulting in a broadened transition. $^4$He also belongs to the 3d-XY universality class and thus 3d-XY scaling should hold. The calculations of Ref. 7 however go beyond the usual 3d-XY scaling approach because they provide an explicit scaling function for the specific heat and the degree of broadening is directly given in terms of the correlation length. A sketch of this finite size scaling applied to superconductors has recently been reported by T. Schneider [8].

In this Letter, using high-resolution thermal expansion data of a YBa$_2$Cu$_3$O$_{7-\delta}$ single crystal in magnetic fields up to 11.4 T, we show that the broadening of the thermodynamic transition in an optimally doped HTSC is identical to that calculated by Haussmann for superfluid $^4$He. (We note that this effect has not been observed in $^4$He because of the large extrinsic rotationally induced broadening due to pressure gradients [7].) This finding has strong implications for the understanding of the H-T phase diagram of HTSC. First, it shows that T$_c$ is essentially a 3d-XY phase ordering transition and that magnetic fields on the order of 10 T only affect the phase coherence. Second, our results suggest that a much higher field scale must exist where the pairing amplitude is reduced.

The thermal expansivity measurements have been performed on a nearly optimally doped (δ=0.05; T$_c$=91.4 ± 0.15 K) detwinned YBa$_2$Cu$_3$O$_{7-\delta}$ single-crystal from the same batch as used in previous studies [9,10]. A high-resolution capacitance dilatometer with a relative resolution of ΔL/L=10$^{-9}$ was used at a constant heating rate of 15 mK/s and in fields up to 11.4 T.

The thermal expansivity is directly proportional to the specific heat close to a phase transition and, as shown previously [9,10], for YBa$_2$Cu$_3$O$_{7-\delta}$ it is of great advantage to examine the expansivity instead of the specific heat due to the much larger superconducting-signal to phonon-background ratio - especially if one considers the difference in the linear expansivities of the a- and b-axes, $\alpha_{b-a}$. FIG. 1a) shows the electronic part of $\alpha_{b-a}$, which is obtained by subtracting the phonon background as shown previously [10]. In a magnetic field the λ-shaped zero-field anomaly, which has been analyzed in detail [9,10], is broadened as it is known from various thermodynamic- and transport measurements [1,3,11-14]. At lower temperatures additional sharp peaks appear in $\alpha_{b-a}$, which are not of electronic origin but are rather due to irreversible magnetostrictive effects at the irreversibility line [15].

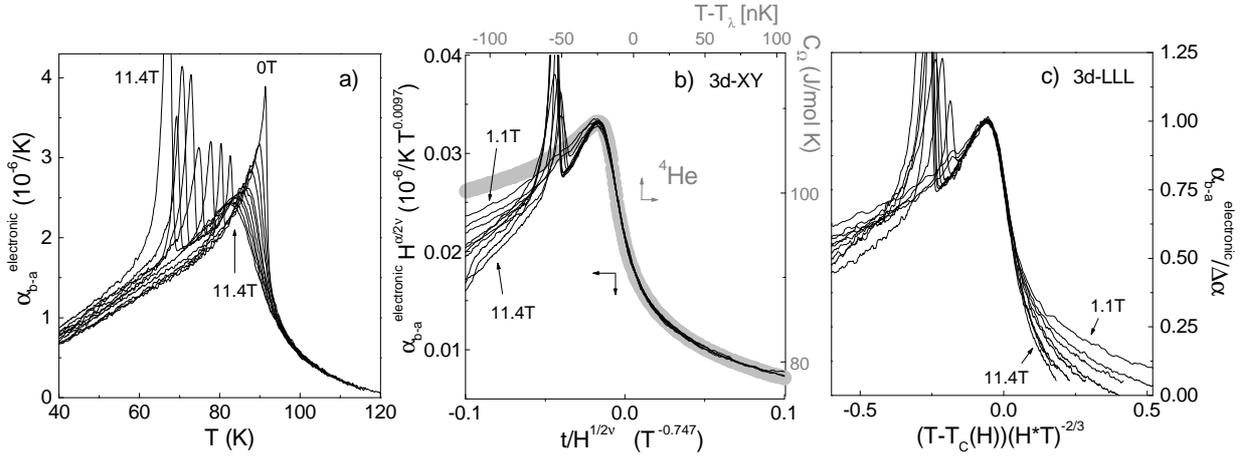

FIG. 1 a) Electronic thermal expansivity, $\alpha_{b-a}^{electronic}$, of YBa$_2$Cu$_3$O$_{6.95}$ in magnetic fields of 0, 1.1, 2.3, 3.4, 4.6, 5.7, 6.8, 8, 9.1, 10.2 and 11.4 T (H ∥ c-axis). b) 3d-XY and c) 3d-LLL scaling of the data in a). The thick gray curve in b) represents the specific heat of rotating (ω=2π/s) $^4$He at the λ-transition from Ref. [7].

The first important result, which directly follows from an inspection of the data in Fig. 1a, is that 11.4 T have no visible effect on the quite sizable fluctuation contribution extending to about 30 K above $T_c$ suggesting that '$T_c$' is not shifted by the field. This is in strong contrast to the behavior of a classical superconductor with fluctuations in which $T_c$, and thus also the temperature at which the fluctuations should diverge, decreases with applied field [16]. The above behavior is, however, expected in the 3d-XY approach; in the following we show that our data exhibit excellent 3d-XY scaling and match the scaling function for the specific heat of $^4$He from Ref. [7].

In Fig. 1b the expansivity data near $T_c$ are plotted using the 3d-XY scaling variables ($\alpha H^{\alpha'/2\nu}$ vs. $t/H^{1/2\nu}$, $t=(T-T_c)/T_c$ with the critical exponents ν=0.669 and α'=-0.013 (not to be confused with the expansivity α). Very good scaling is observed over a wide range of values of $t/H^{1/2\nu}$ extending form the peak in alpha to $t/H^{1/2\nu} = 0.1$ [17]. The appropriately scaled specific heat data for rotating $^4$He (thick gray curve) from Ref. [7] also scales perfectly with our expansivity data suggesting a similar broadening mechanism for both cases. The data do not scale below the peak in the expansivity - we attribute this to the temperature dependence of the 'jump' component, which also affects the scaling in zero field [9]. It is interesting to note that the peaks of the expansivity at the irreversibility temperature also scale quite well suggesting that the irreversibility/ melting transition also obeys 3d-XY scaling, as has been previously [11-13,18].

We also tried to scale our data using the 3d-LLL approach [19,20], in which one plots $\alpha/\Delta\alpha(H)$ vs. $(T-T_c(H))(HT)^{-2/3}$ as shown in Fig. 1c. $\Delta\alpha(H)$ is the field-dependent mean-field component of the anomaly and $T_c(H)$ is the field-dependent mean-field $T_c$, which was chosen to be constant and equal to $T_c(H=0)$. To obtain any kind of scaling close to $T_c$ we have to set $\Delta\alpha(H)$ equal to the anomaly height at the peak. For this case we get reasonably good scaling close to $T_c$. Actually this is not unexpected because close to $T_c$ the 3d-XY and 3d-LLL approaches are mathematically nearly identical and have scaling exponents too similar to be experimentally distinguishable [1]. However, a clear lack of scaling is observed for $(T-T_c(H))(HT)^{-2/3}$ greater than about 0.05. This is not improved by replacing the constant $T_c$ by a $T_c(H)$. The reason we are able for the first time to clearly distinguish between 3d-XY and 3d-LLL scaling is because we are able to check for scaling over a much larger interval of the respective scaling variables due to a better knowledge of the background.

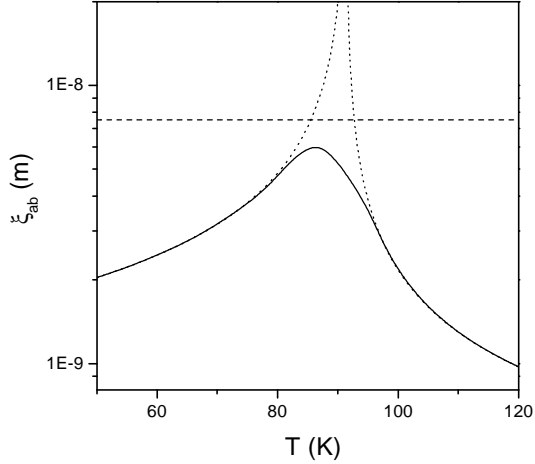

FIG. 2 In-plane correlation length, $\xi_{ab}$, of YBa$_2$Cu$_3$O$_{7-\delta}$ around $T_c$ derived from a comparison with the $^4$He data of Ref. 7. The horizontal dashed line represents the cutoff length scale $\ell$ in a field of 5 T and the dotted line the zero-field divergence.

In Fig. 2 we show the correlation length in the vicinity of $T_c$ obtained by scaling the width of the broadening seen in the thermal expansivity data to the results for $^4$He [7,21]. YBa$_2$Cu$_3$O$_{7-\delta}$, in contrast to $^4$He, is anisotropic, and a field parallel to the c-axis probes the in-plane correlation length, $\xi_{ab}$ [8]. The dotted lines represent the zero-field behavior with $\xi_{ab}^- = 12 \pm 1$ Å and $\xi_{ab}^+ = 4.5 \pm 1$ Å, and the solid line the approximate behavior expected in a field of 5T in analogy to $^4$He [7,22]. As can be seen, the correlation length is cutoff by the length scale $\ell$ (dashed line), which for a superconductor is given by [7,8]

$$\ell = \sqrt{\Phi_o / 2pH}$$
                                                                    - 1 -

In the 3d-XY approach the correlation lengths above, $\xi_{ab}^+$, and below, $\xi_{ab}^-$, $T_c$ follows a power law of the form $x^\pm = x_0^\pm |t|^{-n}$. *Two-scale-factor universality* [22,23] provides an explicit relation between $\xi_{ab}^+$ and $\xi_{ab}^-$ and the specific heat amplitudes $A^+/A^- \cong 1.054$

$$\left(\frac{x_0^-}{x_0^+}\right)^3 = \frac{A^+}{A^-}\left(\frac{R_x^-}{R_x^+}\right)^3 \approx (2.7)^3$$
                                                                    - 2 -

where $R_\xi^- \cong 0.95$ and $R_\xi^+ \cong 0.36$ are universal as well. The ratio $x_0^-/x_0^+ = x^-/x^+ \approx 2.7$, is responsible for the asymmetry in the temperature dependence of $x$ relative to $T_c$. Values of the correlation length can be derived directly from the fluctuation amplitude of the specific heat using Eq. [2], and these values ($\xi_{ab}^- = 13$ Å [23]; $\xi_{ab}^- = 9$ Å [24]) agree very well with ours. This agreement together with the excellent scaling shown in Fig. 1 provides strong evidence that the field-induced broadening in YBaCuO is due to the same 'finite-size' effect as in rotating $^4$He.

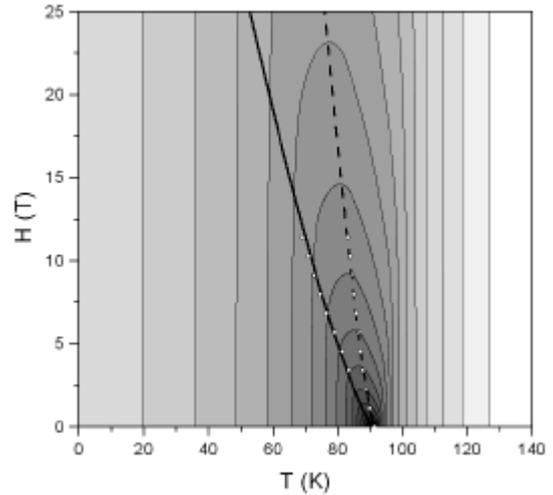

FIG. 3 Contour plot of $\xi_{ab}$ in the H-T plane. Values of $\xi_{ab}$ are represented using a logarithmic gray scale with white ($\log\xi$= -9.2) and black ($\log\xi$= -7.5) separated by steps of 0.1. The solid and dashed lines represent the vortex melting line, $T_m$, and the peak in the expansivity $T_{peak}$, respectively.

Our results are summarized in the H-T phase diagram shown in Fig. 3. The contour lines, which represent constant values of the correlation length $\xi_{ab}$, show virtually no field effect below the vortex melting line (solid line) and above $T_c$(0T) The dashed line represents the peak in the expansivity, $T_{peak}$. Both of these lines nicely follow the expected 3d-XY scaling behavior, in which scaling assures that the ratio $\xi/a$ is constant. Here $a$ is the vortex spacing in an hexagonal lattice given by

$$a = \sqrt{2\Phi_o / \sqrt{3}B} \approx 2.7 \cdot \ell$$
                                                                    - 3 -

$T_{peak}$ occurs near the maximum in the

correlation length and is approximately given by the condition $\xi/a \approx \ell/a \approx 0.37$. This criteria is very similar to the one for classical superconductors at $H_{c2}$ [16]

$$\frac{x_{GL}}{a} = \frac{\sqrt{\Phi_o/2pH_{c2}}}{\sqrt{2\Phi_o/\sqrt{3}H_{c2}}} = 0.371 \ , \qquad -4-$$

which suggests that $T_{peak}$ represents a sensible criteria for '$H_{c2}$' in HTSC and shows that $H_{c2}$ in classical superconductors in a sense is also due to a 'finite-size' effect (overlapping of vortex cores). Physically, $T_{peak}$ corresponds to the broadened remains of the zero-field phase-ordering transition – true phase coherence occurs only at $T_m$. Equations 3 and 4 suggest a close relationship between the correlation and coherence lengths and also provides a justification for the 'classical' determinations of $H_{c2}$ and $\xi$ in HTSC using magnetization measurements ($\xi_{GL} = 16.4$ Å [25], $\xi_{GL} = 10.9$ Å [26] ).

The previous discussion concentrated on the $H_{c2}$ of the 3d-XY phase-ordering transition, $H_{c2}^{3d-XY}$. Our results suggest that a much higher field scale, $H^{pairing}$, might exists where the pairing correlations are destroyed. This is in analogy to the clear separation at zero field of $T_c$ and the pairing at $T^*$ in the preformed-pairs scenario of HTSC. $H_{c2}^{3d-XY}$ and $H^{pairing}$ can thus be viewed as the field dependencies of $T_c$ and $T^*$, respectively. We observe no field effect on the 3d-XY phase-fluctuations above $T_c(0T)$ which demonstrates that fields on the order of 10 T have a negligible effect on the pairing amplitude. A similar conclusion was obtained by recent measurements of the Nernst signal by Y. Wang et al. [27], and their Nernst contour lines are very similar to our Fig. 3. This suggests that $H^{pairing}$ is much larger than $H_{c2}^{3d-XY}$, in agreement with recent experimental [28] and theoretical [29] results, which find that this field scale is on the order of 100-200 T. Finally, in agreement with Ref. [30], we would like to point out that this shows that a negligible magnetic-field effect (for H=10-20 T) on the pseudogap [31,32] does not rule out superconducting pairing correlations as the origin of the pseudogap.

We would like to acknowledge helpful discussions with A. Junod, T. Schneider, G. A. Williams and W. Goldacker.